\documentclass{llncs}
\usepackage{makeidx}  
\usepackage{llncsdoc}
\usepackage{graphicx}
\usepackage{amssymb}
\usepackage{siunitx}  
\usepackage{textcomp}
\usepackage{adjustbox}
\usepackage{comment}
\usepackage{todonotes}
\usepackage{color, soul}
\usepackage{mathtools}

\usepackage{amsmath}

\usepackage{array}
\newcolumntype{L}[1]{>{\raggedright\let\newline\\\arraybackslash\hspace{0pt}}m{#1}}
\newcolumntype{C}[1]{>{\centering\let\newline\\\arraybackslash\hspace{0pt}}m{#1}}
\newcolumntype{R}[1]{>{\raggedleft\let\newline\\\arraybackslash\hspace{0pt}}m{#1}}
\usepackage{placeins}  

\begin{document}
\title{Automated Quality Controlled Analysis of 2D Phase Contrast Cardiovascular Magnetic Resonance Imaging} 

\newcommand*\samethanks[1][\value{footnote}]{\footnotemark[#1]}
\newcommand{\rowstyle}[1]{\gdef\currentrowstyle{#1}%
  #1\ignorespaces
}

\author{Emily Chan \inst{1} \and 
Ciaran O'Hanlon \inst{1} \and 
Carlota Asegurado Marquez  \inst{1} \and
Marwenie Petalcorin  \inst{1} \and
Jorge Mariscal-Harana \inst{1} \and
Haotian Gu \inst{7}
Raymond J. Kim \inst{6} \and
Robert M. Judd \inst{6} \and
Phil Chowienczyk \inst{1, 7}
Julia A. Schnabel \inst{1,2,3} \and
Reza Razavi  \inst{1,4} \and
Andrew P. King \inst{1} \and
Bram Ruijsink* \inst{1,4,5} \and
Esther Puyol-Ant\'on\thanks{Joint last authors.} \inst{1}}
\authorrunning{E Chan et al.}   
\institute{School of Biomedical Engineering \& Imaging Sciences, King\textquotesingle s College London, UK. \and Technical University Munich, Munich, Germany \and Helmholtz Center Munich, Germany \and Guy\textquotesingle{}s and St Thomas\textquotesingle{} NHS Foundation Trust, London, UK \and Department of Cardiology, Heart and Lung Division, University Medical Center Utrecht, Utrecht, The Netherlands. \and Division of Cardiology, Department of Medicine, Duke University, Durham, North Carolina, USA \and British Heart Foundation Centre, King’s College London, London, UK }

\maketitle              

\begin{abstract} 
Flow analysis carried out using phase contrast cardiac magnetic resonance imaging (PC-CMR) enables the quantification of important parameters that are used in the assessment of cardiovascular function. An essential part of this analysis is the identification of the correct CMR views and quality control (QC) to detect artefacts that could affect the flow quantification. We propose a novel deep learning based framework for the fully-automated analysis of flow from full CMR scans that first carries out these view selection and QC steps using two sequential convolutional neural networks, followed by automatic aorta and pulmonary artery segmentation to enable the quantification of key flow parameters. Accuracy values of 0.958 and 0.914 were obtained for view classification and QC, respectively. For segmentation, Dice scores were $>$0.969 and the Bland-Altman plots indicated excellent agreement between manual and automatic peak flow values. In addition, we tested our pipeline on an external validation data set, with results indicating good robustness of the pipeline. This work was carried out using multivendor clinical data consisting of 986 cases, indicating the potential for the use of this pipeline in a clinical setting. \\

\keywordname{cardiac magnetic resonance; deep learning; quality control; cardiac function; view-selection; multivendor}
\end{abstract}

\section{Introduction}
Cardiac Magnetic Resonance (CMR) is one of the most comprehensive modalities for assessment of cardiovascular function, enabling the quantification of cardiac volumes, myocardial motion and tissue characteristics. Aside from these properties, CMR allows for the measurement of blood flow in the major vessels. These measurements can be useful for quantifying a range of aspects of cardiovascular function, including cardiac output, the shunting of blood through intra-cardiac connections, the severity of valvular lesions (including regurgitation), and vascular properties. While 2D flow can potentially be measured in any blood vessel using CMR, pulmonary and aortic flow measurements are the most important measures used in clinical assessments \cite{Nayak2015}. 
After acquisition of 2D Phase Contrast (PC) images, flow measures are normally obtained semi-automatically; a physician manually delineates the target blood vessel and a semi-automatic (pixel threshold based) method propagates this segmentation over the full cardiac cycle. After this, manual adjustment of the target segmentation is performed and, importantly, the physician interrogates the images and obtained flow data for quality. This quality control (QC) includes checking for errors in imaging, such as fold-over and aliasing of the blood flow signal, and checking the appearance of the flow curve for irregular (unphysiological) behaviour \cite{Rebergen1993}. 

Several works have proposed the semi-automatic or automatic quantification of flow using PC-CMR images. For example, Bidhult et al. \cite{Bidhult2019} incorporated shape constraints to aid in active contour based semi-automatic segmentation of the ascending aorta and pulmonary artery. Similarly, Bratt et al. \cite{Bratt2019} utilised a modified U-Net to develop a method for fully-automatic aortic flow quantification. However, these solutions ignore highly important steps in the automation of flow assessment: image classification, to select the correct images from a full CMR scan automatically, and the aforementioned quality control, which is an integral part of CMR (flow) analysis.

In recent years, machine learning-enabled QC of CMR images has received increasing interest, in particular for cine images. Tarroni et al. \cite{Tarroni2019} developed a decision forest-based solution for heart coverage estimation, inter-slice motion detection and image contrast estimation in the cardiac region, while Ruijsink et al. \cite{ruijsink2020fully}  proposed an end-to-end pipeline with automated QC techniques to analyse cine CMR sequences. Other works focused on detecting missing slices \cite{Zhang2016,Zhang2017} and motion artefacts \cite{Oksuz2019}, while some addressed QC of automatic image segmentation \cite{Robinson2019,Alba2018,Hann2019,Wang2020}. Additionally, Vergani et al. \cite{Vergani2021} recently proposed the use of convolutional neural networks for the automatic quality controlled selection of cine images for analysis. To the best of our knowledge, the use of deep learning (DL) for view selection and QC prior to flow quantification has not been explored with PC-CMR images. Moreover, existing solutions for automated flow quantification have generally been trained on single-centre/highly selective data sets, making their generalisation to external data sets challenging.

In this paper, we propose a novel framework that takes into account these important aspects to provide a fully-automated pipeline from image-to-report. We develop a DL-based image classification algorithm to detect 2D PC flow imaging, and classify/select the most important target sequences for analysis (aortic flow and pulmonary flow); a QC algorithm to detect image quality related errors; and a robust segmentation pipeline for 2D pulmonary and aortic flow. Our pipeline calculates key parameters of interest in flow analysis: peak, net, total forward and total backward flow. Furthermore, we utilise an external validation set to assess the generalisability of our framework.

\section{Methods}
\label{sec:methods}
The framework we present is composed of four steps: 1) view selection aimed at identifying conventional PC-CMR views, 2) QC to detect image quality related artefacts, 3) automated segmentation of PC images, and 4) parameter extraction (see Figure \ref{fig:pipeline} for a summary). These steps are described below:

\begin{figure}[ht]
    \centering
    \includegraphics[width=\textwidth]{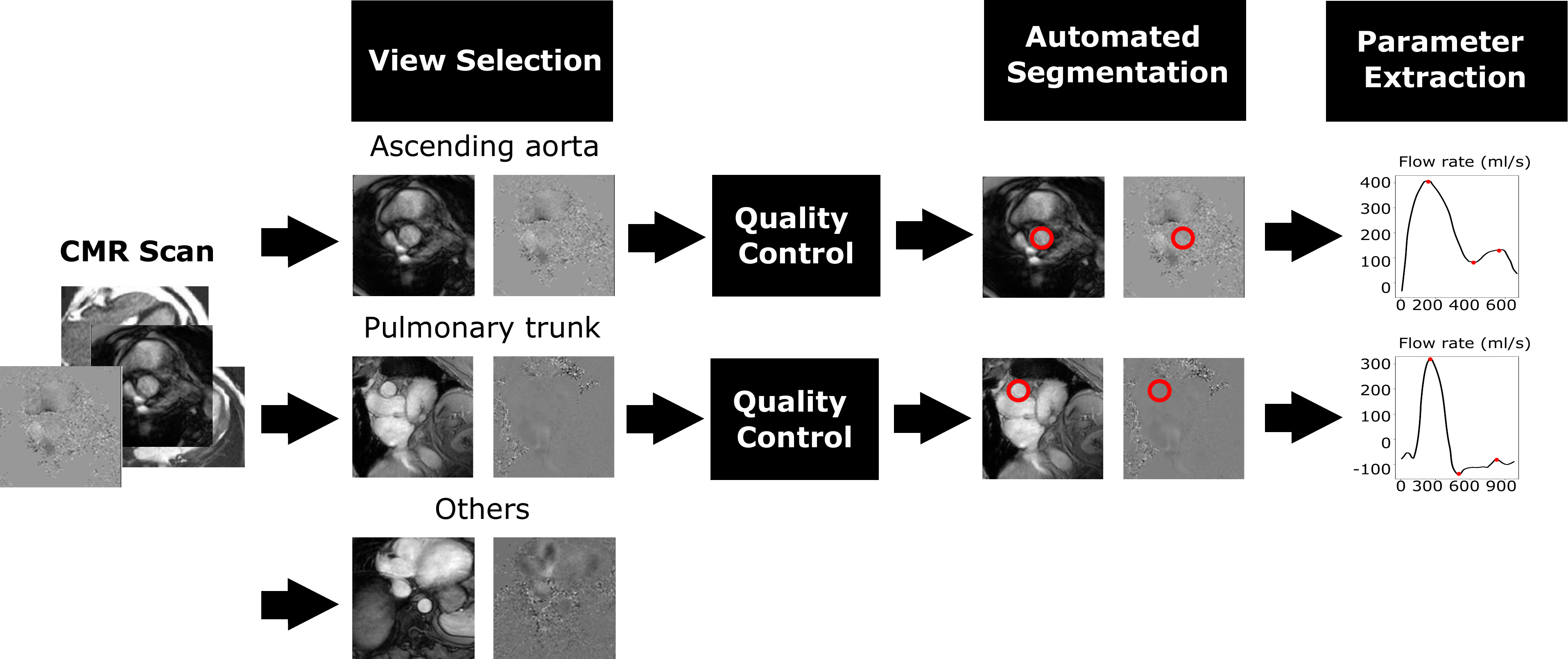}
    \caption{Overview of the proposed AI-based framework for automatic flow quantification from a full CMR scan. The first block, view selection, aims to identify conventional phase contrast views; the second block aims to perform quality control to detect image quality related artefacts; the third block aims to perform automatic segmentation of the quality controlled selected images; and the fourth block aims to extract clinical parameters using the resulting segmentations.}
    \label{fig:pipeline}
\end{figure}

\textbf{1. View Selection}: The first part of our framework aims to identify the standard 2D PC views used for the analysis of cardiac function and haemodynamics from a CMR acquisition. The manually classified data were divided as follows: 80\% was used for training, 10\% was used for validation, and 10\% for testing. The training, validation and test data cohorts had a mutually exclusive subject pool, i.e. acquisitions from the same subject could only be used in one of the three cohorts. 
Prior to training, all images were cropped to a standard size of 192x192 pixels. We used the AlexNet network \cite{Krizhevsky2012} to classify between the ascending aorta, pulmonary trunk and other views. The network was trained with batch size 16 and learning rate 0.001 with stochastic gradient descent for 200 epochs with cross entropy loss to classify images into the three classes described. For the training data, data augmentation was performed on-the-fly using random translations (±30 pixels), rotations (±90°), flips (50\% probability) and scaling (up to 20\%) for each mini-batch of images before feeding them to the network. The probability of augmentation for each of the parameters was 50\%.

\textbf{2. Quality Control}: QC of image acquisition is of particular importance when measuring flow because errors in the quantification of net flow can arise from a range of different image acquisition or processing quality issues, including tortuosity of the aorta, planning of the acquisition plane, arrhythmia and breathing artefacts. These quality issues can lead to errors in flow analysis and therefore need to be detected. We implemented the VGG-11 \cite{Simonyan2015} network as a binary classifier to detect such artefacts, utilising manual labels provided by a cardiologist as ground truth. The same training, validation and test splits of the data utilised for view selection were used here. Similarly, the same pre-processing and data augmentation scheme were carried out for this step in the framework, with the same network parameters used for training, although binary cross entropy loss was used to classify the images as those with and without artefacts.

\textbf{3. Automated Segmentation}: We used the ‘nnU-Net’ framework \cite{isensee2021nnu} to segment all frames through the cardiac cycle for the ascending aorta and pulmonary artery
classes, respectively. We used the 2D ‘nnU-Net’ implementation and the models were trained for 1000 epochs. Training was performed with a batch size of 32 using stochastic gradient descent with Nesterov momentum ($\mu$=0.99) and an initial learning rate of 0.01. The loss function was the sum of the cross entropy and Dice losses. Data augmentation was performed on the fly and included rotations, scaling, Gaussian noise, Gaussian blur, brightness, contrast, simulation of low resolution, gamma correction and mirroring. An ensemble of five different models trained on subsets of the training data was used to achieve the final predicted segmentations on the test set. All deep learning models used for this work were trained on an NVIDIA Quadro RTX 6000 GPU.

 

\textbf{4. Flow Curve Estimation \& Parameter Estimation:} pixel values were converted to velocity using the established formula \cite{watanabe2019accuracy} as follows:
\begin{equation*}
    \text{Velocity} = \frac{10 \pi R}{\text{VENC}} *P*M
\end{equation*}
where $\text{P}$ and $\text{M}$  represent the raw pixel values from the phase and magnitude maps, $\text{VENC}$ is an adjustable scanner parameter representing the maximum measurable flow velocity extracted and R is a reconstruction scaling factor specified from the PC-CMR DICOM files.

Flow was calculated from the automatic segmentation map of a given PC scan as:
\begin{equation*}
    \text{Net Flow} = \sum_{n=0}^{N} \sum_{i=0}^{I}  S_{n,i} V_{n,i} a \Delta_t
\end{equation*}
%
where $n$ is the frame index, $N$ the number of temporal frames in the scan, $I$ the number of pixels in each frame, $S$ the binary segmentation map, $V$ the velocity map calculated using the previous equation, $a$ is the pixel area (in cm\textsuperscript{2}), and $\Delta_t$ the time interval between frames.

From the flow curves, we quantify the peak, total forward, total backward and net flow.


\section{Materials}
\label{sec:materials}
This is a retrospective multivendor study conducted on a large set of clinical 2D PC-CMR data from three centres that includes the full spectrum of cardiac disease phenotypes. Details of the databases are provided below:\\
\textbf{1. UK Biobank (UKBB):} This database contains a mixture of 80 patients and healthy volunteers. CMR imaging was performed using a 1.5 T Siemens MAGNETOM Aera (see \cite{petersen2015uk} for further details of the image acquisition protocol).\\
\textbf{2. Guy’s and St Thomas’ NHS Foundation Trust (GSTFT):} This database contains 830 patients acquired under routine clinical CMR practice with a range of different cadiovascular diseases (CVDs). CMR imaging was performed using different CMR scanners (Philips Achieva 1.5T/3.0T, Philips Ingenia 1.5T and Siemens Aera 1.5T).\\
\textbf{3. Duke University Hospital (Duke):} This database contains 102 patients acquired in routine clinical CMR practice with a range of different CVDs. CMR imaging was performed using different CMR scanners (Siemens Avanto 1.5T, Siemens Sola 1.5T, Siemens Verio 3.0T and Siemens Vida 3.0T). This database is used as an external validation data set.\\

For the three clinical databases, all data were manually classified and annotated though the full cardiac cycle by multiple cardiologists, and an additional review of all segmentations was performed to ensure the annotations were of high quality.
Table \ref{table:1} summarises the cases used for view classification, segmentation and QC.

\begin{table}[ht] 
\caption{Number of images used in this study for view selection, automated segmentation and quality control (QC). Ao: ascending aorta, PA: pulmonary artery.}
\centering
\begin{tabular}{l | ccc | c |cc} \hline
 & \multicolumn{3}{c}{\textbf{View Selection}} & \multicolumn{1}{|c}{\textbf{QC}} & \multicolumn{2}{|c}{\textbf{Segmentation}}\\ 
Database & Ao & PA & Other & Ao and PA & Ao & PA \\ \hline
UKBB & 80 & 0 & 0 & 80 &  79 & 0 \\
GSTFT & 284 & 215 & 18 & 499 &  477 & 353 \\
Duke & 54 & 48 & 0 & 100 &  40 & 36 \\ \hline
\end{tabular}
\label{table:1}
\end{table}

\section{Experiments and Results}
\label{sec:results} 
We present the results in terms of the internal data set; this comprises the UKBB and GSTFT data sets, which were merged and then randomly split into the training/validation/test splits described in Section \ref{sec:methods}. External data set is used to refer to the Anon Centre 2 data set, which was used for external validation, where 20\% of this data was combined with the original training cohort and used to finetune the network before testing on the remaining external validation data.

\textbf{1. View Selection}:
The performance of the view selection network was evaluated in terms of precision, recall and the F1-score for each of the three classes, as well as the accuracy achieved for all classes together. Table \ref{table:view_classification} summarises these results for the internal data set and the external data set. High performance is shown on the internal data set, with an accuracy of 0.958, while the accuracy of 0.833 on the external data set demonstrates good generalisability for the view selection task.

\textbf{2. Quality Control:}
As with the view classification network, the performance of the quality control network was measured using the precision, recall, F1-score and accuracy, which are reported in Table \ref{table:QC}. The results demonstrate good performance on both the internal and external data sets, with accuracy values of 0.914 and 0.850, respectively.

\begin{table}[ht] 
\caption{View classification results, in terms of precision, recall and F1 score for each class, as well as the global accuracy.}
\centering
\begin{tabular}{l | ccc | ccc} \hline
 & \multicolumn{3}{c}{\textbf{Internal Data Set}} & \multicolumn{3}{|c}{\textbf{External Data Set}}\\  
 & Precision & Recall & F1-score & Precision & Recall & F1-score \\ \hline
Ascending Aorta  & 1.000 & 0.945 & 0.972 & 0.785 & 0.944 & 0.857 \\
Pulmonary Artery & 0.913 & 0.977 & 0.944 & 0.912 & 0.708 & 0.800 \\
Other            & 0.800 & 1.000 & 0.889 & - &  - & - \\ \hline
Accuracy         & \multicolumn{3}{c}{0.958} & \multicolumn{3}{|c}{0.833} \\ \hline
\end{tabular}
\label{table:view_classification}
\end{table}

\begin{table}[ht] 
\caption{Quality control performance, in terms of the precision, recall, F1-score and accuracy.}
\centering
\begin{tabular}{l | c | c |c | c}
\hline
 & \textbf{Precision} & \textbf{Recall} & \textbf{F1-score} & \textbf{Accuracy}\\ \hline
Internal Data Set & 0.914 & 0.913 & 0.907 & 0.914 \\
External Data Set & 0.873 & 0.850 & 0.792 & 0.850 \\ \hline
\end{tabular}
\label{table:QC}
\end{table}

\textbf{3. Segmentation:}
The segmentation network for each view was evaluated in terms of the average Dice coefficient between the manual and automatic segmentations, and the results for the internal and external validation sets can be seen in Table \ref{table:seg}. Dice scores for the internal data set were excellent, with good values also obtained for the external data set, where values of greater than 0.969 and greater than 0.800 were achieved, respectively.
\begin{table}[ht] 
\caption{Segmentation results in terms of the mean Dice coefficient for the two segmentation networks (ascending aorta and pulmonary artery) for the full cardiac cycle.}
\centering
\begin{tabular}{l | c | c} \hline
& \textbf{Internal Data Set} & \textbf{External Data Set}\\  \hline
Ascending Aorta  & 0.969 & 0.871 \\
Pulmonary Artery & 0.971 & 0.800 \\ \hline
\end{tabular}
\label{table:seg}
\end{table}

\begin{figure}[htb]
    \vspace{-3pt}
    \begin{minipage}[b]{0.49\linewidth}
      \centering
      \centerline{\includegraphics[width=\linewidth]{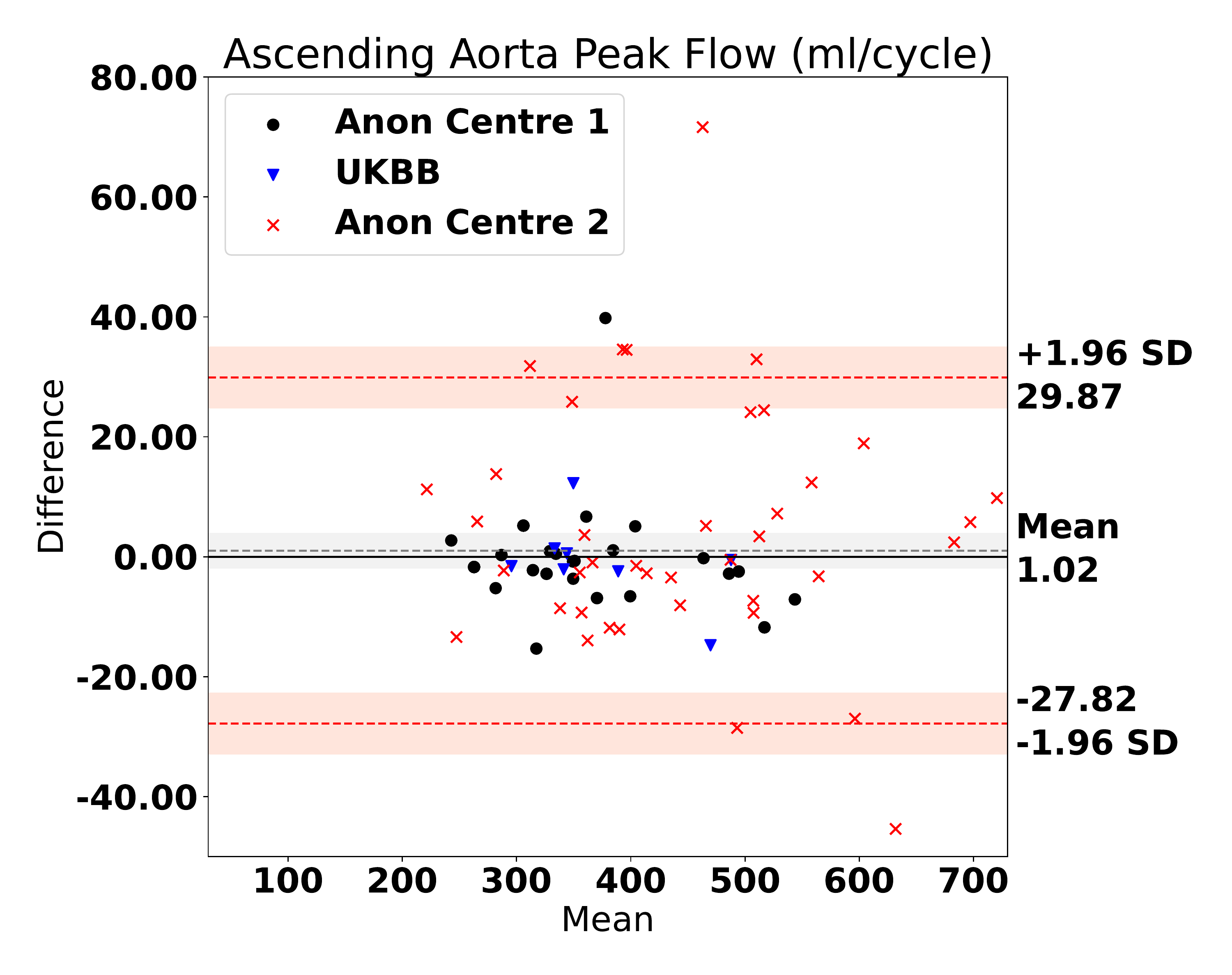}}
    \end{minipage}
    \vspace{-7pt}
    \hfill
    \begin{minipage}[b]{0.49\linewidth}
      \centering
      \centerline{\includegraphics[width=\linewidth]{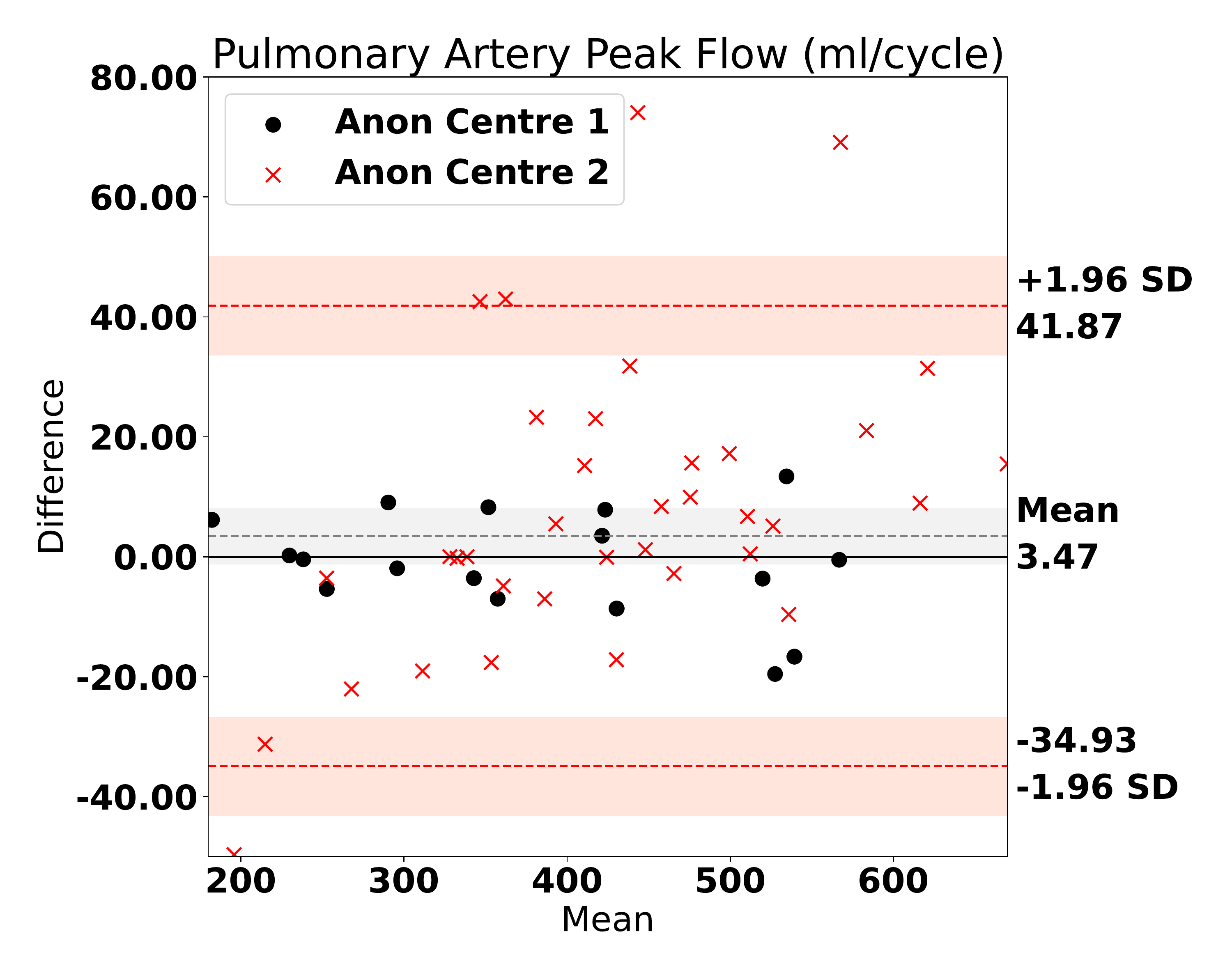}}
    \end{minipage}
    \begin{minipage}[b]{0.49\linewidth}
      \centering
      \centerline{\includegraphics[width=\linewidth]{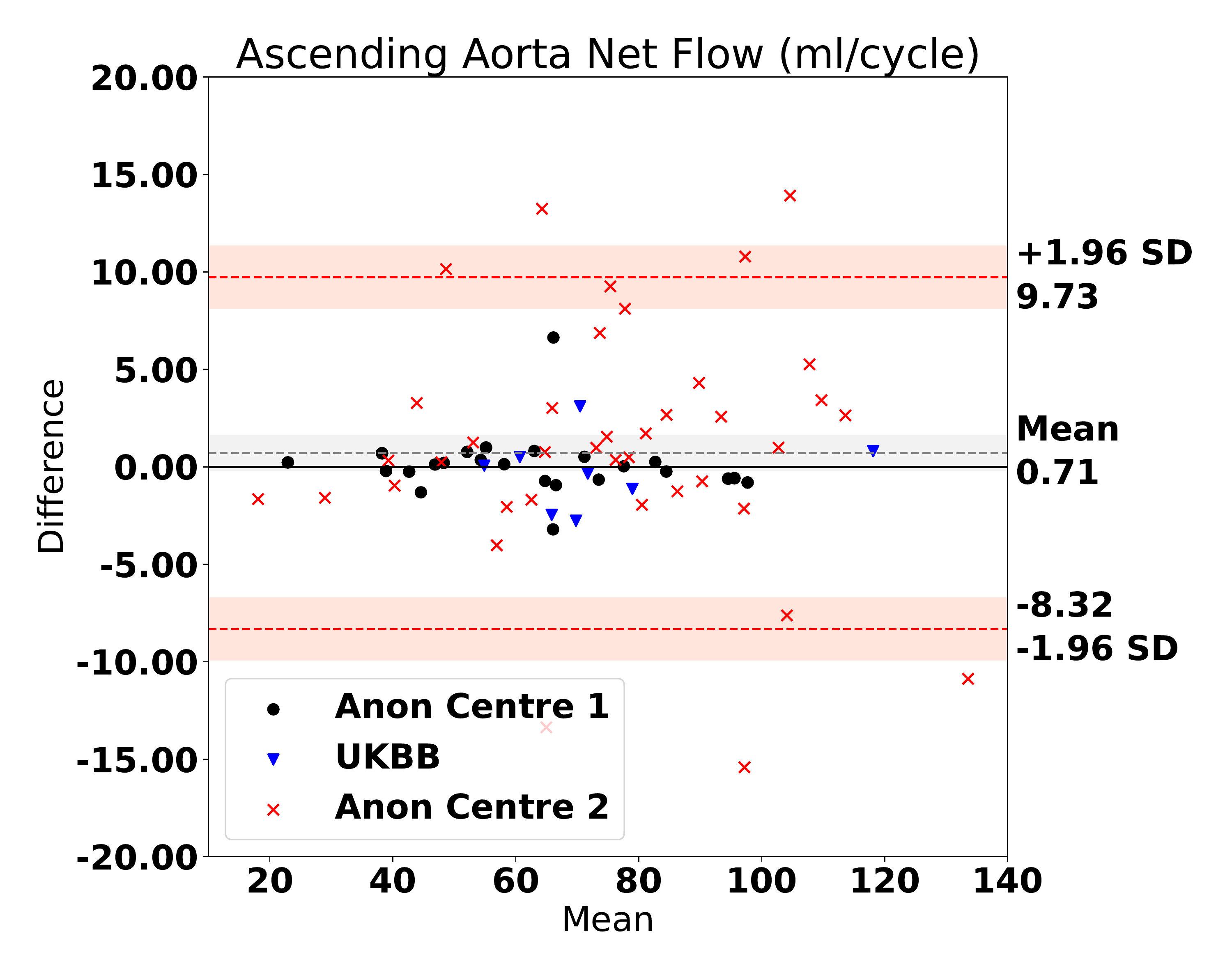}}
    \end{minipage}
    \hfill
    \vspace{-7pt}
    \begin{minipage}[b]{0.49\linewidth}
      \centering
      \centerline{\includegraphics[width=\linewidth]{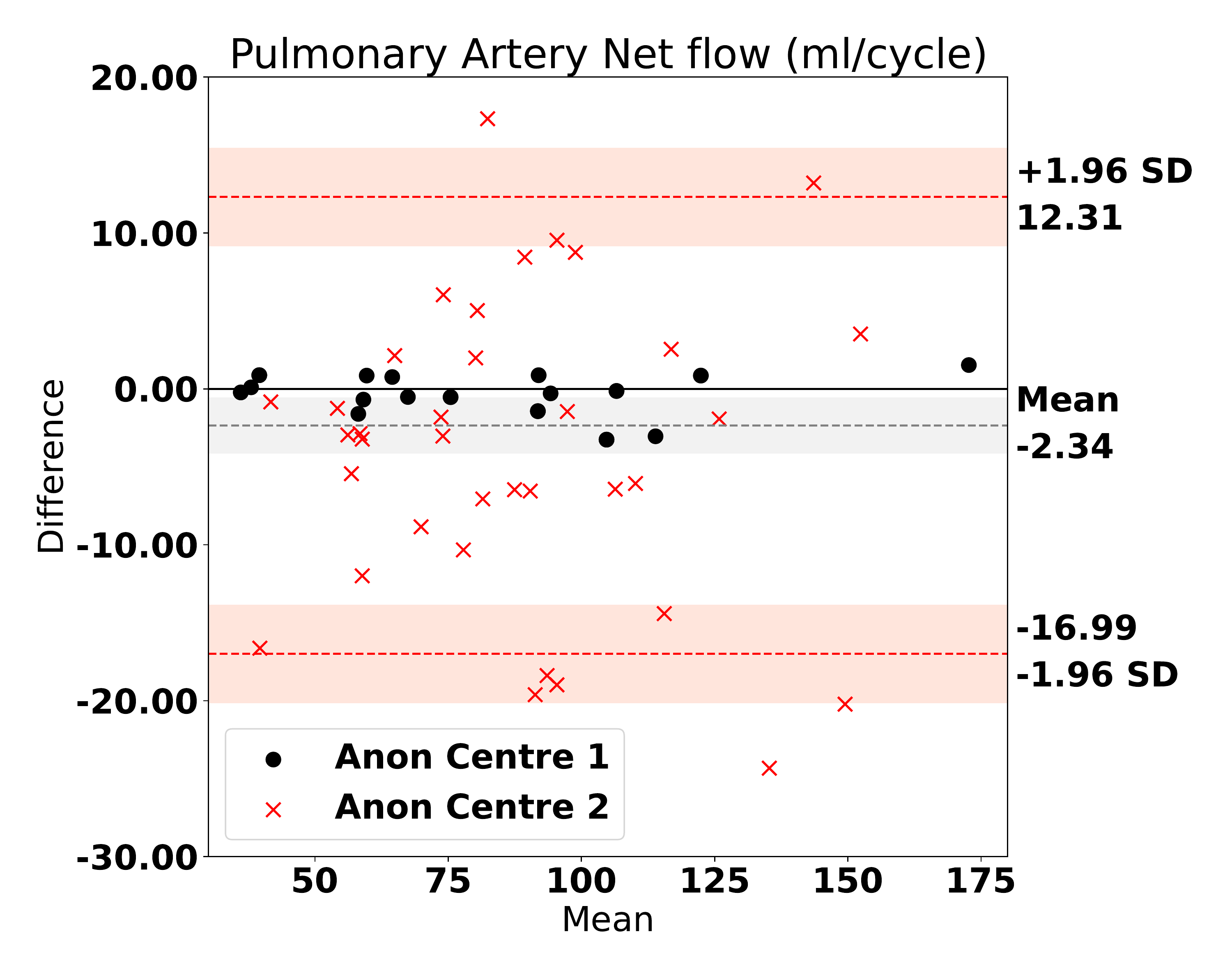}}
    \end{minipage}
    \begin{minipage}[b]{0.49\linewidth}
      \centering
      \centerline{\includegraphics[width=\linewidth]{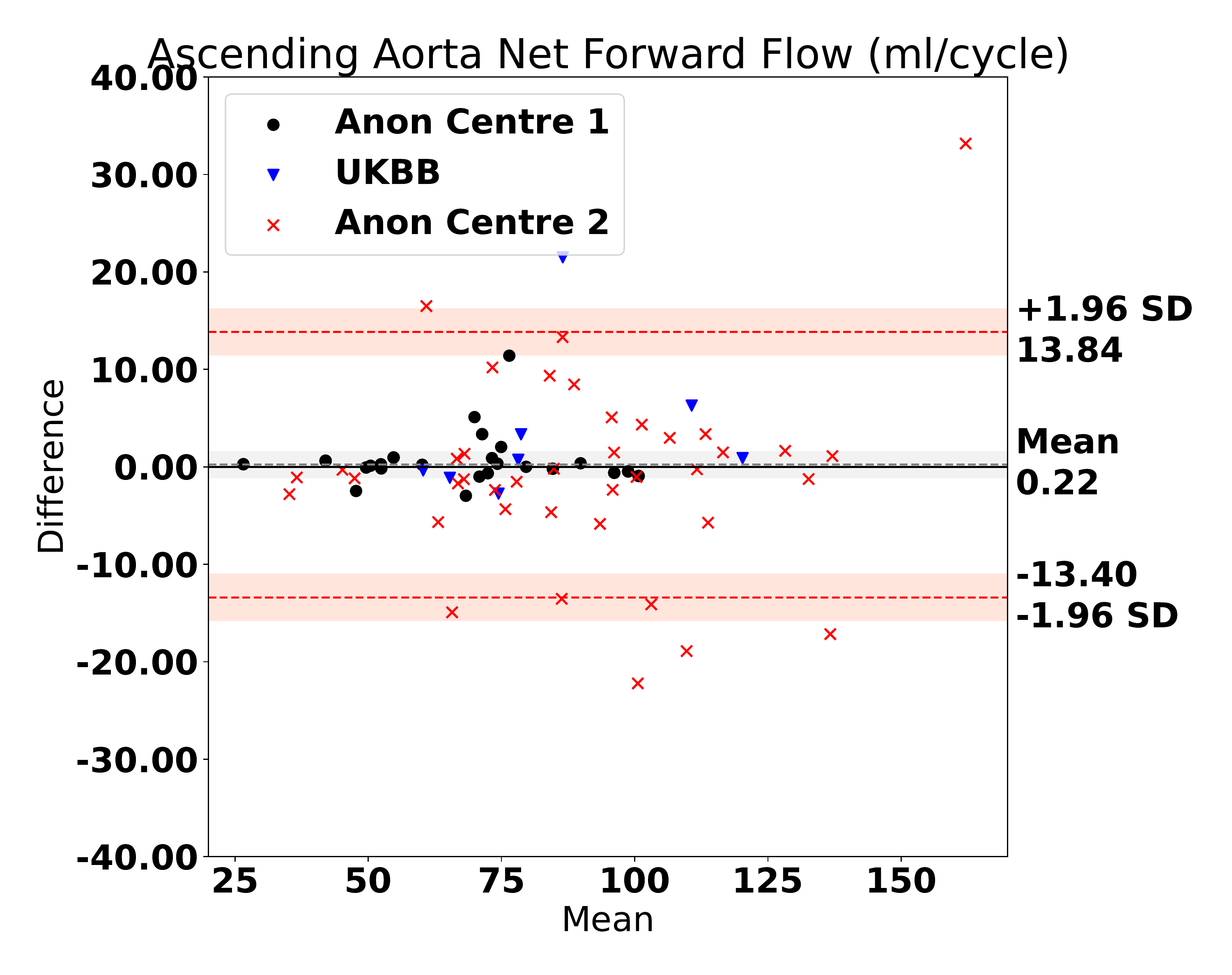}}
    \end{minipage}
    \hfill
    \vspace{-7pt}
    \begin{minipage}[b]{0.49\linewidth}
      \centering
      \centerline{\includegraphics[width=\linewidth]{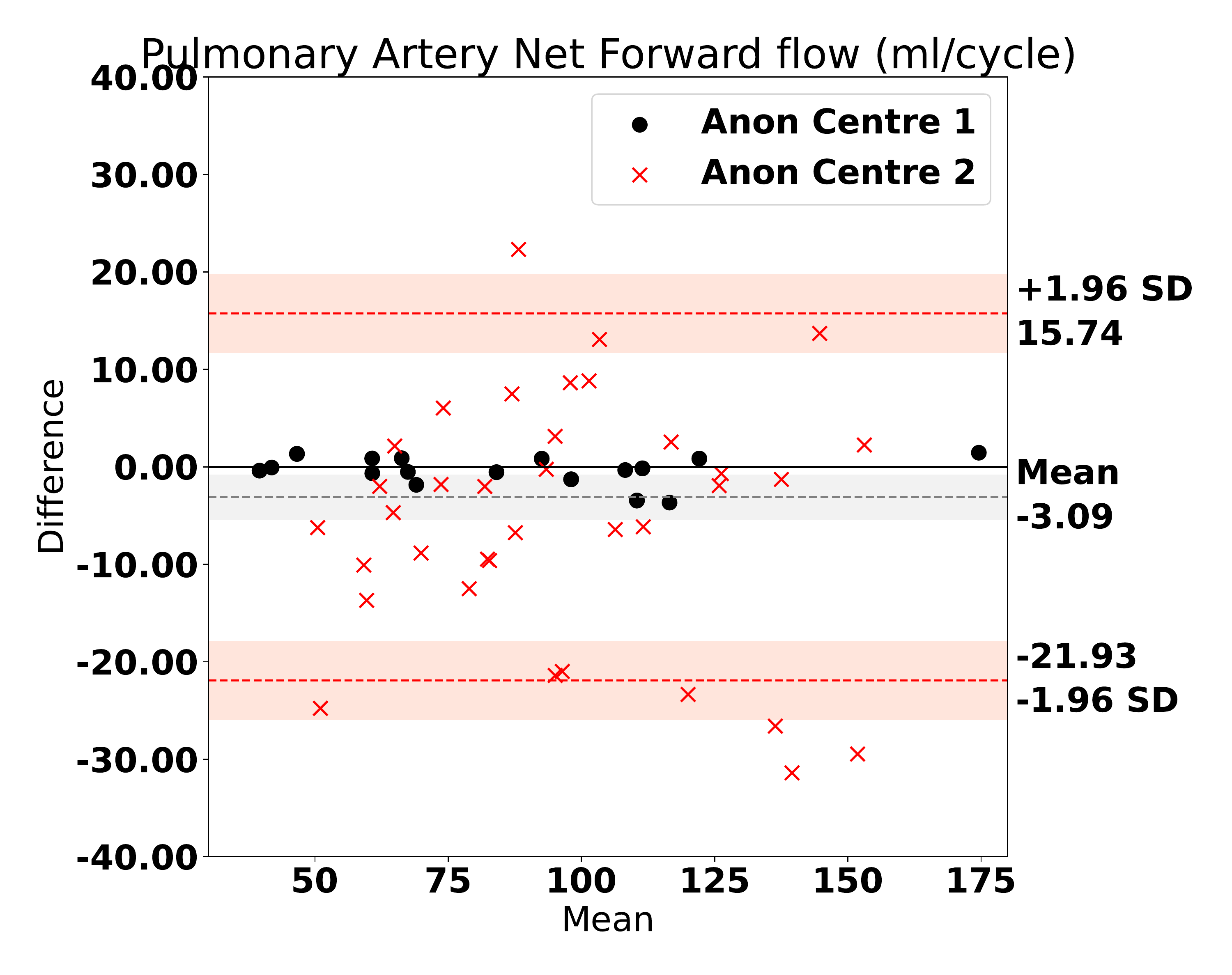}}
    \end{minipage}
    \begin{minipage}[b]{0.49\linewidth}
      \centering
      \centerline{\includegraphics[width=\linewidth]{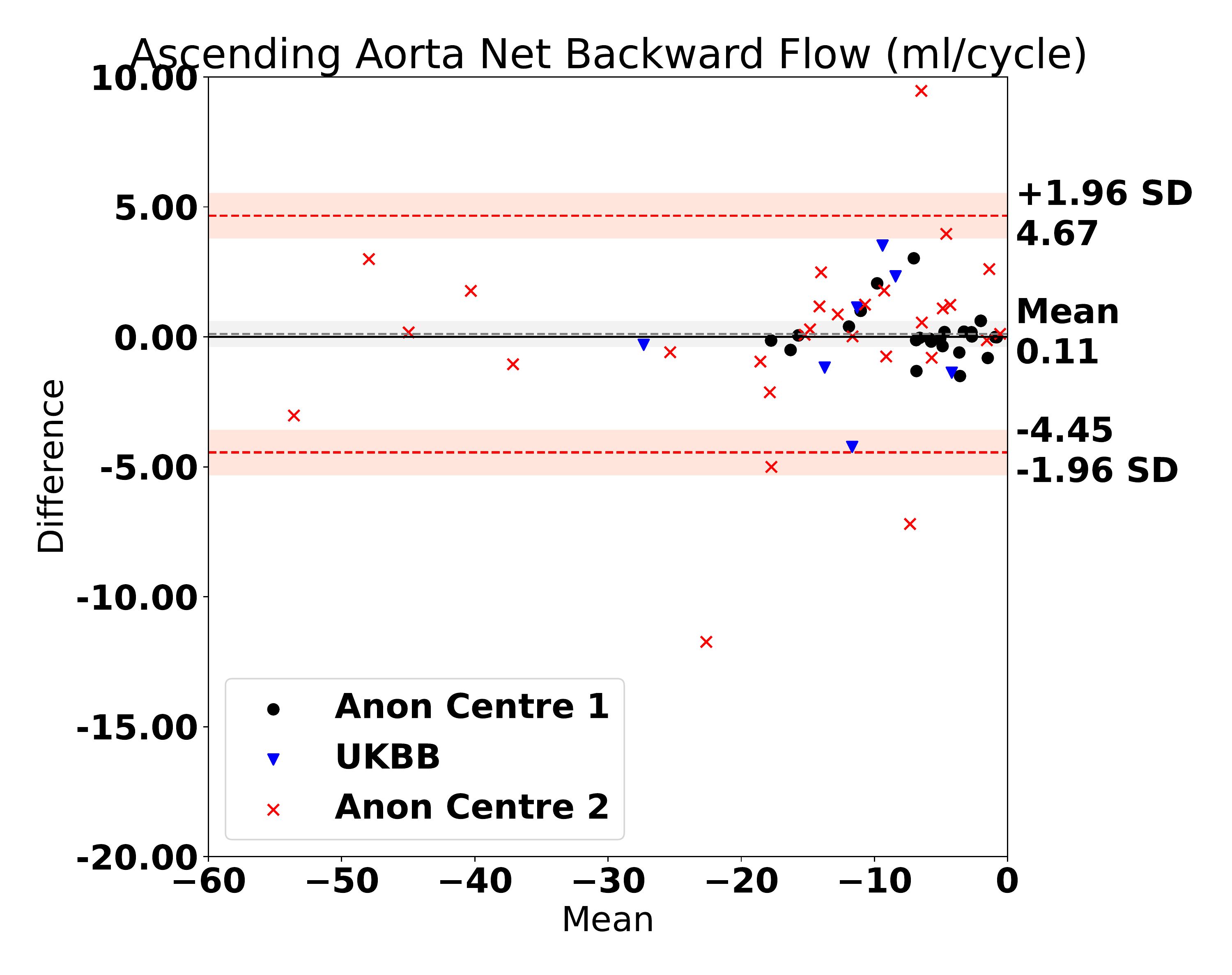}}
    \end{minipage}
    \vspace{-7pt}
    \hfill
    \begin{minipage}[b]{0.49\linewidth}
      \centering
      \centerline{\includegraphics[width=\linewidth]{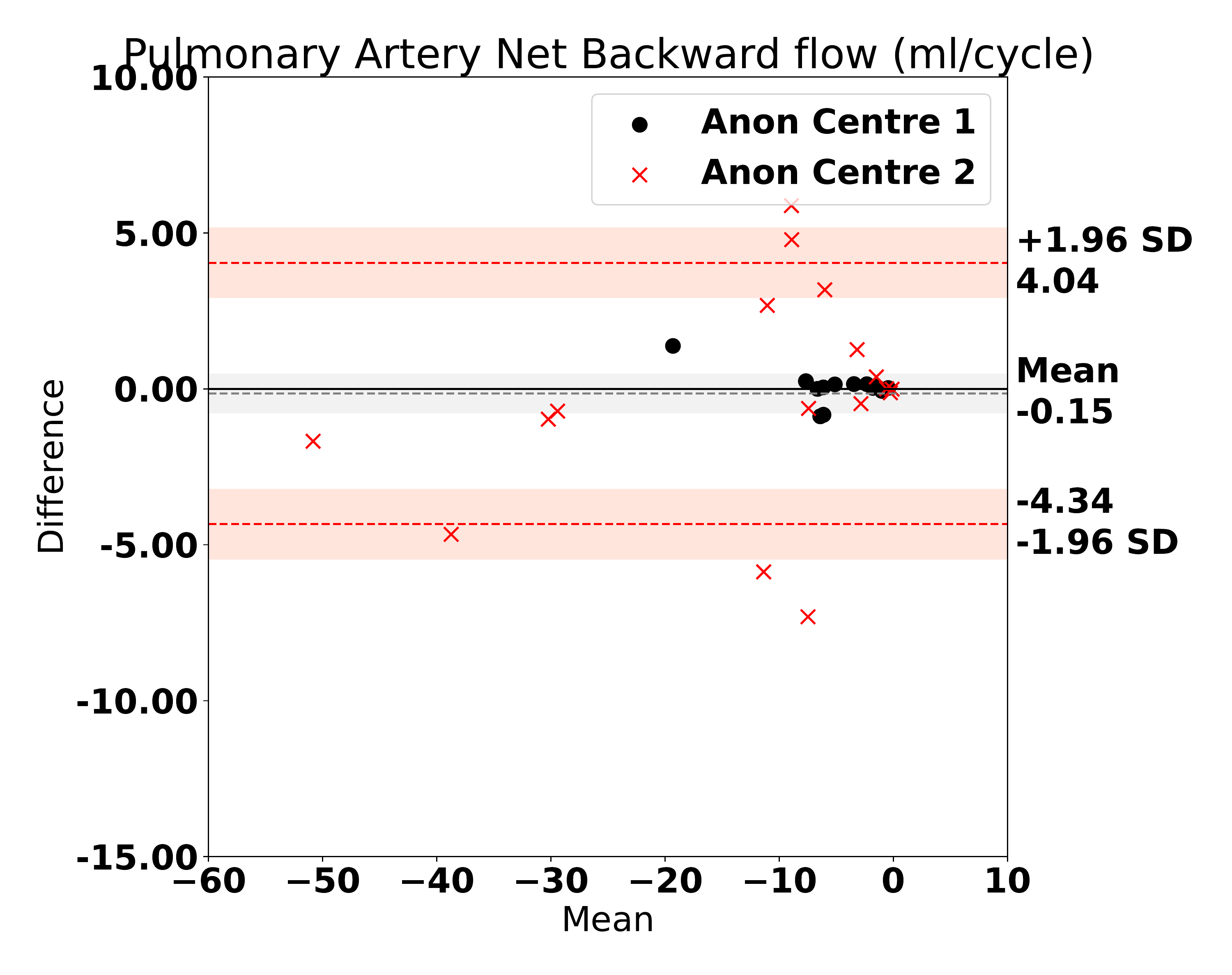}}
    \end{minipage}
    \vspace{-7pt}
    \caption{Bland-Altman plots for peak flow (first row), net flow (second row), forward flow (third row) and backward flow(final row), for the ascending aorta (left column) and the pulmonary artery (right column).}
    \label{fig:BA_plots}
\end{figure}

\textbf{4. Parameter Estimation:}
The peak, net, total forward and total backward flow values in the aorta and pulmonary artery estimated using the automatic segmentations were compared to the manually obtained values using Pearson's correlation and Bland-Altman plots. Table \ref{table:corr} shows the correlation coefficients; for all views, correlations are strong, with coefficients $>$0.918 (p$<$0.001). Figure \ref{fig:BA_plots} shows the generated Bland-Altman plots; there is good agreement between automated and manual analysis, without a significant bias for any of the views and narrow limits of agreement. 

\begin{table}[ht] 
\caption{Pearson correlation coefficient values obtained when comparing the manual and automatic values for peak, net, total forward and total backward flow (p$<$0.001).}
\centering
\begin{tabular}{l | cccc | cccc} \hline
& \multicolumn{4}{c}{\textbf{Internal Data Set}} & \multicolumn{4}{|c}{\textbf{External Data Set}} \\
& Peak & Net & Forward & Backward & Peak & Net & Forward & Backward\\ \hline
Ascending Aorta  & 0.997 & 0.998 & 0.988 & 0.980 & 0.998 & 0.970 & 0.953 & 0.918\\
Pulmonary Artery & 0.998 & 0.999 & 0.999 & 0.993 & 0.986 & 0.956 & 0.930 & 0.939\\ \hline
\end{tabular}
\label{table:corr}
\end{table}

\vspace{-5mm}
\section{Discussion}
\label{sec:discussion}

In this work, we have proposed a novel end-to-end pipeline enabling automatic flow analysis from full CMR scans, which encompasses four main steps: view selection to detect conventional 2D PC-CMR scans; quality control to identify images with artefacts; automatic segmentation of the ascending aorta and pulmonary artery views; and the estimation of key flow parameters. To the best of our knowledge, this is the first automatic flow analysis framework to include the automatic quality-controlled selection of conventional 2D PC-CMR views, which mimics crucial steps ordinarily carried out by physicians. We utilised a large scale database from multiple centres, vendors and field strengths, with one data set being used for external validation. 

Our results show a good performance at each step in the pipeline, with accuracy values of 0.958 and 0.914 for view classification and QC, respectively, and Dice scores $>$0.969, indicating that our pipeline achieves close to human-level performance and is comparable to other automatic flow quantification frameworks \cite{Bidhult2019,Bratt2019}. The obtained flow curves can be used to calculate further metrics of interest in cardiovascular medicine, including ejection times and first-phase ejection fraction (EF1). Additionally, we obtained good results on the external validation set, indicating that our pipeline has good generalisability. This is noteworthy, as the external data set included images that were obtained with a different imaging strategy (higher image-acceleration factors and breath-held imaging), leading to significant differences in image characteristics.

A limitation of our pipeline is that the output automatic segmentations are not checked for quality before being used for the computation of the flow parameters. Future work will therefore include the addition of an extra quality control step after the segmentation of the vessels to ensure the quality of the parameter extraction. Furthermore, in this work, we excluded views other than the ascending aorta and pulmonary artery from analysis. In the future, we aim to extend our pipeline to these other views, such as descending aorta, branch pulmonary artery and pulmonary vein flows.

\section*{Acknowledgements}
This work was supported by the UKRI London Medical Imaging \& Artificial Intelligence Centre for Value Based Healthcare, and the Wellcome EPSRC Centre for Medical Engineering at the School of Biomedical Engineering and Imaging Sciences, King\textquotesingle s College London (WT 203148/Z/16/Z). The authors also acknowledge financial support from the National Institute for Health Research (NIHR) Cardiovascular MedTech Co-operative award to the Guy’s and St Thomas’ NHS Foundation Trust and the Department of Health National Institute for Health Research (NIHR) comprehensive Biomedical Research Centre award to Guy’s \& St Thomas’ NHS Foundation Trust in partnership with King’s College London.

\FloatBarrier
\bibliographystyle{splncs03}
\bibliography{refs.bib}

\end{document}